\documentclass{aastex}
\usepackage{emulateapj5}

\newcommand{\kms}{km~s$^{-1}$}

\slugcomment{To appear in ApJ (2002 Dec 10 issue)} 

\shorttitle{SN 1988Z}
\shortauthors{Williams et al.}

\begin{document}

\title{Radio Emission from SN 1988Z and Very Massive Star Evolution} 

\author{Christopher L. Williams}
\affil{P.O.~Box 16148, Stanford University, Stanford CA 94309; clmw@stanford.edu}

\author{Nino Panagia\altaffilmark{1}}
\affil{Space Telescope Science Institute, 3700 San Martin Drive, Baltimore, MD 21218; panagia@stsci.edu}

\author{Schuyler D. Van Dyk}
\affil{IPAC/Caltech, Mail Code 100-22, Pasadena, CA 91125; vandyk@ipac.caltech.edu}

\author{Christina K. Lacey}
\affil{Univ.~of South Carolina, Dept. of Physics \& Astronomy, Columbia, SC 29208; lacey@sc.edu}

\author{Kurt W. Weiler}
\affil{Naval Research Laboratory, Code 7213, Washington, DC 20375-5320;kurt.weiler@nrl.navy.mil}

\and

\author{Richard A. Sramek}
\affil{National Radio Astronomy Observatory, P.O.~Box 0, Socorro, NM 87801; dsramek@nrao.edu}

\altaffiltext{1}{On assignment from the Astrophysics Division, Space Science Department of ESA.}

\begin{abstract}
We present observations of the radio emission from the unusual supernova SN 1988Z in MCG +03-28-022 made with the Very Large Array at
20, 6, 3.6, and 2 cm, including new observations from 1989 December 21, 385 
days after the optically
estimated explosion date, through 2001 January 25, 4,438 days after
explosion. At a redshift $z = 0.022$ for the
parent galaxy ($\sim$100 Mpc for $\rm{H}_0 = 65 ~\rm{km} ~\rm{s}^{-1}
~\rm{Mpc}^{-1})$, SN 1988Z is the most distant radio supernova ever
detected. With a 6 cm maximum flux density of 1.8 mJy, SN 1988Z is
$\sim$20\% more luminous than the unusually powerful radio supernova
SN 1986J in NGC 891 and only $\sim$3 times less radio luminous at 6 cm
peak than the extraordinary SN 1998bw, the presumed counterpart to GRB
980425. Our analysis and model fitting of the radio light curves for
SN 1988Z indicate that it can be well-described by a model involving
the supernova blastwave interacting with a high-density circumstellar
cocoon, which consists almost entirely of clumps or filaments.  SN
1988Z is unusual, however, in that around age 1750 days the flux
density begins to decline much more rapidly than expected from the
model fit to the early data, without a change in the absorption
parameters.  We interpret this steepening of the radio flux density
decline rate as due to a change in the number density of the clumps in
the circumstellar material (CSM) without a change in the average properties
of a clump.  If one assumes that the blastwave is traveling through
the CSM at $\sim2,000$ times faster than the CSM was established
(20,000 \kms ~vs. 10 \kms), then this steepening of the emission
decline rate represents a change in the presupernova stellar wind
properties $\sim10,000$~yrs before explosion, a characteristic time
scale also seen in other radio supernovae.  Further analysis of the
radio light curves for SN 1988Z implies that the SN progenitor star
likely had a ZAMS mass of $\sim20$--30 $M_\odot$.  We propose
that SNe, such as SN 1986J, SN 1988Z, and SN 1998bw, with very 
massive star progenitors 
and associated
massive wind ($\dot M\gtrsim 10^{-4}\ M_\odot$ yr$^{-1}$)
have very highly-clumped, wind-established CSM and unusually high
blastwave velocities ($>20,000$~\kms).  

\end{abstract}

\keywords{supernovae, individual (SN 1988Z), galaxies, individual (MCG 
+03-28-022, Zw 095-049), radio continuum, stellar evolution: massive stars}

\section{Introduction}

Supernova
(SN) 1988Z was independently discovered in MCG +03-28-022 (Zw 095-049) 
near $\rm{m_B} \sim 16.4$ by both G.~Candeo on 1988 December 12 
(Cappellaro, Turatto, \& Candeo 1988) and C.~Pollas on 1988 December 14 
(Pollas 1988). The SN was classified as a Type II, based on the detection of 
hydrogen emission lines in the optical spectra by Cowley \& Hartwich 
(Heathcote et al.~1988), which also indicated that the SN was quite distant, 
with redshift $z \sim 0.022$.  Filippenko (1989) confirmed the Type II 
identification, with possible resemblance to SN 1987F. The spectra revealed at 
early times that SN 1988Z was peculiar, with a remarkably blue color, a lack 
of absorption lines and P-Cygni profiles, very narrow ($\lesssim$100 km 
s$^{-1}$ FWHM) [O III] circumstellar emission lines, and a steadily
growing, relatively narrow ($\simeq$2000 km s$^{-1}$ FWHM) component to the H 
I and He I lines (Stathakis \& Sadler 1991; Filippenko 1991a,b). 
Stathakis \& Sadler (1991), and Turatto et al.~(1993) for later times, 
analyzed in detail the spectroscopic and photometric observations of SN 1988Z.
Schlegel (1990) proposed that SN 1988Z, along with other similar SNe,
constituted
a new class of SNe, the Type II-``narrow,'' or SNe IIn.

The properties of the optical spectra and light curves indicated strong SN 
shock-circumstellar shell
interaction (Filippenko 1991a,b; Turatto et al.~1993) which, together with its
resemblance optically
to SN 1986J (Filippenko 1991a,b; see also Rupen et al.~1987,
Leibundgut et al.~1991), strongly suggested that SN 1988Z should be a very 
luminous radio emitter, as SN 1986J is a strong radio source (see Weiler,
Panagia, \& Sramek 1990).
Sramek et al.~(1990) reported the radio detection at 6 cm wavelength of SN 1988Z 
with the Very Large Array (VLA)\footnote{The VLA telescope of the National Radio 
Astronomy Observatory is operated by Associated Universities, Inc. under a 
cooperative agreement with the National Science Foundation.} 
on 1989 December 21. The supernova is located at 
RA(J2000) = $10^h 51^m 50\fs138$, 
Dec(J2000) = $+16\arcdeg 00\arcmin 00\farcs16$,
with an uncertainty of $\pm$ 0\farcs2 in each coordinate, 
which is coincident, to within the uncertainties, with its optical position 
(Kirshner, Leibundgut, \& Smith 1989). 
This initial announcement of the detection of radio emission was followed 
by a 
more thorough study and analysis of the first three years of multifrequency 
radio observations by Van Dyk et al.~(1993a).  They compared SN 1988Z in 
detail with SN 1986J and concluded that the two SNe are very similar in their 
radio properties.  
They suggested that the progenitor to SN 
1988Z was a massive [$20 \leq~ M(M_\odot) \leq 30$] star which underwent a 
high mass-loss phase ($\dot M \ga 10^4\ M_\odot$ yr$^{-1}$) 
before explosion (see also Stathakis \& Sadler 1991).

Luminous radio emission has also been an indicator of X-ray emission from
SNe, with synchrotron radio emission being produced as the forward SN shock
interacts with the CSM and X-rays being emitted from the corresponding reverse
shock region interacting with the SN ejecta (Chevalier \& Fransson 1994).  For instance,
SN 1986J was detected as a luminous X-ray source by Bregman \& Pildis (1992)
and Houck et al.~(1998).
Fabian \& Terlevich (1996) reported the detection of X-rays from SN 1988Z with
ROSAT.
SN 1988Z is a very luminous X-ray emitter with a bolometric X-ray
luminosity of $\sim 10^{41}~\rm{erg}~\rm{s}^{-1}$.

Chugai \& Danziger (1994) 
offered two models to explain the unusual characteristics of SN 1988Z: 1)
shock interaction with a two component wind consisting of a tenuous,
homogeneous medium with embedded higher density clumps, or 2) shock
interaction with a similar tenuous, homogeneous medium and a higher-density,
equatorial, wind-established, disk-like component, favoring
the former over the latter.
However, they unexpectedly
conclude that the SN ejecta is of low mass 
($M < 1\ M_\odot$), and that SN 1988Z may have originated 
from a relatively low-mass star of 
$M_{\rm ZAMS} \sim 8$--10 $M_{\odot}$, in  
sharp contrast to the high-mass 
progenitor suggested by Van Dyk et al.~(1993a) and 
Stathakis \& Sadler (1991).

Aretxaga et al.~(1999) collect 
the observations from X-ray to radio and
attempt to estimate the integrated electromagnetic energy
radiated by SN 1988Z in the first 8.5 years after discovery. They obtain
a value of $\ge 2 \times 10^{51}$ erg, perhaps as high as
$10^{52}$ erg, which they consider is sufficiently high to suggest that SN
1988Z could be classified as a ``hypernova,'' approaching the 
2--$5 \times 10^{52}$ ergs estimated to have been released in SN
1998bw, the possible counterpart of GRB 980425 (Iwamoto et al.~1998), and 
perhaps
indicative of the collapse of the stellar progenitor core into a black hole.
(It is interesting to note here that two SNe IIn, 1997cy and
1999E, may also have been associated with $\gamma$-ray bursts; see 
Pastorello et al.~2002 and references therein).
Aretxaga et al.~(1999) suggest an ejecta mass 
of $\sim15\ M_\odot$, and, therefore, a very high-mass progenitor.

Obviously, SN 1988Z is an extremely interesting and, in many ways,
unusual object.  Fortunately, due to its intrinsic brightness it has
been relatively well-studied in many
wavelength bands.  Here we report radio observations of SN 
1988Z
at multiple wavelengths, including new observations which add another six 
years of monitoring, more than doubling the coverage reported by
Van Dyk et al.~(1993a).  We further interpret its radio emission using a clumpy
wind model.  We conclude that SN 1988Z, like SNe 1986J and, possibly,
other SNe IIn, as well as the unusual
SN 1998bw, arise from the explosions of very massive stars surrounded by highly filamentary CSM.

\section{Radio Observations}

New radio observations of SN 1988Z have been made with the VLA at 20 cm
(1.425 GHz), 6 cm (4.860 GHz), 3.6 cm (8.440 GHz) and 2.0 cm (14.940 GHz) from 
1993 May 4 through 2001 January 25 and are presented in Table 1 and Figure 1.  
The previously published results from Van Dyk et al.~(1993a), along
with a few additional and 
reanalyzed flux density values, are also included 
in Table 1 and Figure 1 for ease of reference.  

Note that the estimated explosion date of 
1988 January 23 obtained by Van Dyk et al.~(1993a) from their model fit was 
very poorly constrained by their fitting procedure.  Here, we adopt the
explosion date of 1988 December 1 estimated from optical data 
(Stathakis \& Sadler 1991).  We have correspondingly modified the SN age at each of the
epochs in Table 1 and Figure 1 relative to those reported in Van Dyk et 
al.~(1993a).  This change in assumed explosion date, although 
large, is not critical for the model fitting and does not significantly 
affect the quality of the fit or the conclusions made by Van Dyk et al.~(1993a).

The techniques of observation, editing, calibration, and error estimation
are described in previous publications on the radio emission from SNe
(see, e.g., Weiler et al.~1986, 1990, 1991).  The ``primary'' calibrator was 
3C~286, which is assumed to be constant in time with flux densities of 14.45, 
7.42, 5.20, and 3.45 Jy at 20, 6, 3.6, and 2 cm, respectively.  The ``secondary''
calibrators\footnote{Secondary calibrators are chosen to be compact and unresolved 
by the longest VLA baselines.  While compact and serving as good phase references, 
such objects are usually variable, so that their flux density must be recalibrated 
from the primary calibrators for each observing session.} were J1051+213, used through 
1991 January 17, and J1125+261, for all epochs after that, with defined positions of 
RA(J2000) = $10^{\rm h}51^{\rm m}48{\fs}789$, 
Dec(J2000) =   +$21^{\circ}19\arcmin 52{\farcs}31$ and 
RA(J2000) = $11^{\rm h}25^{\rm m}53{\fs}712$, 
Dec(J2000) = +$26^{\circ}10\arcmin 19{\farcs}98$, respectively. 
After flux density calibration by 3C~286, they served as the actual gain and phase 
calibrators for SN 1988Z.  As expected for secondary calibrators, the flux densities 
of J1051+213 and J1125+261 have been varying over the years, as can be seen in Table 2 
and Figure 2.

The flux density measurement errors for SN 1988Z are a combination of the rms map
error, which measures the contribution of small unresolved fluctuations
in the background emission and random map fluctuations due to receiver
noise, and a basic fractional error $\epsilon$, included to account for
the normal inaccuracy of VLA flux density calibration (see, e.g.,
Weiler et al.~1986) and possible deviations of
the primary calibrator from an absolute flux density scale.  The final
errors ($\sigma_f$) given for the measurements of SN 1988Z are taken as

\begin{equation}
\label{eq1}
\sigma_{f}^{2} = (\epsilon S_0)^2+\sigma_{0}^2 
\label{eq:err}
\end{equation} where
$S_0$ is the measured flux density, $\sigma_0$ is the map rms for each
observation, and $\epsilon =0.05$ for 20, 6, and 3.6 cm, and $\epsilon
= 0.075$ for 2 cm. 

\section{Parameterized Radio Light Curves}

Following Weiler et al.~(1986, 1990) and Montes, Weiler \& Panagia~(1997), we adopt 
a parameterized model (N.B.: The notation is extended and rationalized here 
from previous publications.  However, the ``old'' notation of $\tau$, 
$\tau^{\prime}$, and $\tau^{\prime \prime}$, which has been used previously, 
is noted, where appropriate, for continuity.):

\begin{eqnarray}
\label{eq2}
S(\mbox{mJy}) = K_1 \left(\frac{\nu}{\mbox{5\
GHz}}\right)^{\alpha} \left(\frac{t-t_0}{\mbox{1\ day}}\right)^{\beta}
e^{-\tau_{\rm external}} \nonumber \\ \times \left(\frac{1-e^{-\tau_{{\rm CSM}_{\rm clumps}}}}{\tau_{{\rm CSM}_{\rm clumps}}}\right) \left(\frac{1-e^{-\tau_{\rm internal}}}{\tau_{\rm internal}}\right) , 
\end{eqnarray} 

\noindent with  

\begin{equation}
\label{eq3}
\tau_{\rm external}  =  \tau_{{\rm CSM}_{\rm homogeneous}}+\tau_{\rm distant} = \tau + \tau^{\prime\prime},
\end{equation}

\noindent where

\begin{equation}
\label{eq4}
\tau_{{\rm CSM}_{\rm homogeneous}} = \tau  =  K_2
\left(\frac{\nu}{\mbox{5 GHz}}\right)^{-2.1}
\left(\frac{t-t_0}{\mbox{1\ day}}\right)^{\delta} ,
\end{equation}

\begin{equation}
\label{eq5}
\tau_{\rm distant}  =   \tau^{\prime\prime}  =  K_4  \left(\frac{\nu}{\mbox{5\
GHz}}\right)^{-2.1} ,
\end{equation} 

\noindent and

\begin{equation}
\label{eq6}
\tau_{{\rm CSM}_{\rm clumps}}  =   \tau^{\prime} =  K_3 \left(\frac{\nu}{\mbox{5\
GHz}}\right)^{-2.1} \left(\frac{t-t_0}{\mbox{1\
day}}\right)^{\delta^{\prime}} ,
\end{equation} 

\noindent with $K_1$, $K_2$, $K_3$, and $K_4$ corresponding, formally,
to the flux density ($K_1$), homogeneous ($K_2$, $K_4$), and clumpy or
filamentary ($K_3$) absorption at 5~GHz one day after the explosion
date, $t_0$.  The terms $\tau_{{\rm CSM}_{\rm homogeneous}}$ and
$\tau_{{\rm CSM}_{\rm clumps}}$ describe the attenuation of local,
homogeneous CSM and clumpy CSM that are near enough to the SN progenitor
that they are altered by the rapidly expanding SN blastwave.
$\tau_{{\rm CSM}_{\rm homogeneous}}$ is produced by an ionized medium that
homogeneously covers the emitting source (``homogeneous external
absorption''), and the $(1-e^{-\tau_{{\rm CSM}_{\rm clumps}}})
\tau_{{\rm CSM}_{\rm clumps}}^{-1}$ term describes the attenuation
produced by an inhomogeneous medium (``clumpy absorption''; see
Natta \& Panagia 1984 for a more detailed discussion of attenuation in
inhomogeneous media).  Both terms have a radial dependence which,
under a constant mass-loss rate, constant velocity wind assumption, is
r$^{-2}$ (see, e.g., Van Dyk et al.~1994 for an example where, for SN
1993J, the radial dependence of the CSM density is found to be flatter
than r$^{-2}$).  The values of $\delta$ and
$\delta^{\prime}$ determine the actual radial density profile if a
constant shock velocity is assumed.  The $\tau_{\rm distant}$ term 
describes the attenuation produced by a homogeneous medium which completely 
covers the source, but is so far from the SN progenitor that it is not 
affected by the expanding SN blastwave and is constant in time.  All absorbing 
media are assumed to be purely thermal, singly ionized gas, which absorbs via 
free-free (f-f) transitions with frequency dependence $\nu^{-2.1}$ in the radio.  
The parameters $\delta$ and $\delta'$ describe the time dependence of the 
optical depths for the local homogeneous and clumpy or filamentary media, 
respectively. 

The f-f optical depth outside the emitting region is proportional to the 
integral of the square of the CSM density over the radius.  Since in the 
simple model (Chevalier 1982a,b) the CSM density decreases as $r^{-2}$, the 
external optical depth will be proportional to $r^{-3}$, and since the radius 
increases as a power of time, $r \propto t^m$ (with $m \leq 1$; i.e., $m = 1$ 
for undecelerated blastwave expansion), it follows that the deceleration parameter, 
$m$, is

\begin{equation}
\label{eq7}
m = -\delta / 3.
\end{equation}

\noindent The Chevalier model relates $\beta$ and $\delta$ to the energy spectrum 
of the relativistic particles $\gamma$ ($\gamma = 2\alpha-1$) by 
$\delta = \alpha - \beta - 3$, so that, for cases where $K_2 = 0$ and $\delta$ is, 
therefore, indeterminate, we can use

\begin{equation}
\label{eq8}
m = -(\alpha - \beta - 3)/3.
\end{equation}

Since it is physically realistic and may be needed in some RSNe where radio 
observations have been obtained at early times and high frequencies, we have also 
included in equation (\ref{eq2}) the possibility for an internal absorption 
term\footnote{Note that, for simplicity, we use an internal absorber attenuation 
of the form $\left(\frac{1-e^{-\tau_{{\rm CSM}_{\rm internal}}}}{\tau_{{\rm CSM}_{\rm internal}}}\right)$, 
which is appropriate for a plane-parallel geometry, instead of the more complicated 
expression (e.g., Osterbrock 1974) valid for the spherical case.  The assumption does 
not affect the quality of our analysis because, to within 5\% accuracy, the optical 
depth obtained with the spherical case formula is simply three-fourths of that 
obtained with the plane-parallel slab formula.}.  This internal absorption  
($\tau_{\rm internal}$) term may consist of two parts -- synchrotron self-absorption 
(SSA; $\tau_{{\rm internal}_{\rm SSA}}$), and mixed, thermal f-f absorption/non-thermal 
emission ($\tau_{{\rm internal}_{\rm ff}}$), so that  

\begin{equation}
\label{eq9}
\tau_{\rm internal}  = \tau_{\rm internal_{\rm SSA}} + \tau_{\rm internal_{\rm ff}} ,
\end{equation}

\noindent where

\begin{equation}
\label{eq10}
\tau_{\rm internal_{\rm SSA}} = K_5\left(\frac{\nu}{\mbox{5\
GHz}}\right)^{\alpha-2.5}  \left(\frac{t-t_0}{\mbox{1\
day}}\right)^{\delta^{\prime\prime}}
\end{equation}

\noindent and

\begin{equation}
\label{eq11}
\tau_{\rm internal_{\rm ff}}  =   K_6  \left(\frac{\nu}{\mbox{5\
GHz}}\right)^{-2.1} \left(\frac{t-t_0}{\mbox{1\
day}}\right)^{\delta^{\prime\prime\prime}} ,
\end{equation}

\noindent with $K_5$ corresponding, formally, to the internal, non-thermal ($\nu^{\alpha - 2.5}$) 
SSA and $K_6$ corresponding, formally, to the internal thermal ($\nu^{-2.1}$) f-f absorption 
mixed with nonthermal emission at 5~GHz one day after the explosion date, $t_0$.  The parameters 
$\delta^{\prime\prime}$ and $\delta^{\prime\prime\prime}$ describe the time dependence of the 
optical depths for the SSA and f-f internal absorption components, respectively. 

A cartoon of the expected structure of an SN and its surrounding media is presented by 
Weiler, Panagia, \& Montes (2001; their Figure 1).  The radio emission is expected to arise 
near the blastwave (Chevalier \& Fransson 1994).

The success of the basic parameterization and modeling has been shown by the good correspondence 
between the model fits and the data for all subtypes of RSNe, e.g., the Type Ib SN 1983N 
(Sramek, Panagia, \& Weiler 1984), the Type Ic SN 1990B (Van Dyk et al.~1993b), the Type II 
SNe 1979C (Weiler et al.~1991, 1992a; Montes et al.~2000) and 1980K 
(Weiler et al.~1992b, Montes et al.~1998), and SN 1988Z (Van Dyk et al.~1993a).  
(Note that, after day $\sim4000$, the evolution of the radio emission from both SNe 1979C and 1980K
deviates from the expected model evolution, and SN 1979C shows a sinusoidal modulation in 
its flux density prior to day $\sim$4000.)  A more detailed discussion of SN radio observations 
and of modeling results is given in Weiler et al.~(2002).
 
Thus, the radio emission from SNe appears to be relatively well understood in terms of blastwave 
interaction with a structured CSM, as described by the Chevalier (1982a,b) model and its modifications
by Weiler et al.~(1986, 1990) and Montes et al.~(1997).  For example, the fact that the homogeneous 
external absorption exponent $\delta \sim -3$, or somewhat less, for most RSNe is evidence that 
the absorbing medium is generally a wind with density $\rho \propto r^{-2}$, as expected from a 
massive stellar progenitor which explodes during the red supergiant (RSG) phase.  

Additionally, in their study of the radio emission from SN 1986J, Weiler et al.~(1990) found that 
the simple Chevalier (1982a,b) model could not describe the relatively slow turn-on.  They therefore 
included terms described mathematically by $\tau_{{\rm CSM}_{\rm clumps}}$ in equations (\ref{eq2}) 
and (\ref{eq6}).  This modification greatly improved the quality of the fit and was interpreted 
by Weiler et al.~(1990) to represent the possible presence of filaments or clumps in the CSM.  
Such a clumpiness in the wind material was again required for modeling the radio data from 
SN 1993J (Van Dyk et al.~1994) and SN 1988Z (Van Dyk et al.~1993a).
Since that time, evidence for filamentation in the envelopes of SNe has also been found from 
optical and UV observations (e.g., Filippenko, Matheson, \& Barth 1994: Spyromilio 1994).

Through use of this modeling a number of physical properties of SNe can be determined from 
the radio observations.  One of these is the mass-loss rate from the SN progenitor prior to
explosion.
From the Chevalier (1982a,b) model, the turn-on of the radio emission for RSNe provides a 
measure of the presupernova mass-loss rate to wind velocity ratio ($\dot M/w_{\rm wind}$).  
Weiler et al.~(1986) derived this ratio for the case of pure, external absorption by a homogeneous 
medium.  However, we now recognize that several possible origins for absorption exist and 
generalize equation (16) of Weiler et al.~(1986) to

\begin{eqnarray}
\label{eq12}
\frac{\dot M (M_\odot\ {\rm yr}^{-1})}{( w_{\rm wind} / 10\ {\rm km\ s}^{-1} )} = 
3.0 \times 10^{-6}\ <\tau_{{\rm eff}}^{0.5}> \ m^{-1.5} \times \nonumber \\ {\left(\frac{v_{\rm i}}{10^{4}\ {\rm km\ s}^{-1}}\right)}^{1.5} \times {\left(\frac{t_{\rm i}}{45\ {\rm days}}\right) }^{1.5} {\left(\frac{t}{t_{\rm i}}\right) }^{1.5 m}{\left(\frac{T}{10^{4}\ {\rm K}} \right)}^{0.68} .
\end{eqnarray}
\\
\noindent Since the appearance of optical lines for measuring SN ejecta velocities 
is often delayed relative to the time of explosion, we normally adopt $t_{\rm i}$ = 45 days.  
Because our observations generally have shown that $0.8 \le m \le 1.0$, and from equation 
(\ref{eq12}) $\dot M \propto t_{\rm i}^{1.5(1-m)}$, the dependence of the calculated mass-loss 
rate on the date $t_{\rm i}$ of the initial ejecta velocity measurement is weak, 
$\dot M \propto t_{\rm i}^{<0.3}$.  Thus, we generally adopt the best optical or VLBI velocity 
measurements available, without worrying about the deviation of their exact measurement epoch 
from the assumed 45 days after explosion.  
For SN 1988Z we follow Turatto et al.~(1993) and assume $v_{\rm i} = 20,000$ \kms\ from the 
highest velocity measured for the broad components of the emission lines.  One has to keep in 
mind that the actual shock velocity for SN 1988Z may be even higher, because it is very hard 
to derive accurate velocities from fitting very broad lines where the extreme wings may be 
lost in the continuum.  

We also normally adopt values of $T = 20,000$ K, $w_{\rm wind} = 10$ \kms\ (which is appropriate 
for a RSG wind), $t = (t_{\rm 6\ cm\ peak} - t_0)$ days from our best fits to the radio data,
and $m$ from equation (\ref{eq7}) or (\ref{eq8}), as appropriate.  The optical depth absorption 
term, $<\tau_{{\rm eff}}^{0.5}>$, however, is not as simple as that used by Weiler et al.~(1986).

Weiler et al.~(2001) were able to identify at least three possible absorption regimes: 
1) absorption by a homogeneous external medium, 2) absorption by a clumpy or filamentary external 
medium with a statistically large number of clumps, and 3) absorption by a clumpy or filamentary 
medium with a statistically small number of clumps.  Each of the three cases requires a 
different formulation for $<\tau_{{\rm eff}}^{0.5}>$.  From our consideration of the radio 
light curves for SN 1988Z, we conclude that the Case 3 of Weiler et
al.~(2001, 2002) is most appropriate (see $\S4$).

Case 3 occurs when the clump number density is small, the probability that the line-of-sight 
from a given clump intersects another clump is low, and both emission and absorption effectively 
occur within each clump.  Such a situation will yield a range of optical depths, from zero for 
clumps on the far side of the blastwave-CSM interaction region, to a maximum corresponding to 
the optical depth through a clump for clumps on the near side of the blastwave-CSM interaction 
region.  We expect an attenuation of the form, 
$({1-e^{-\tau_{{\rm CSM}_{\rm clumps}}}}){\tau_{{\rm CSM}_{\rm clumps}}^{-1}}$, as given in 
equation (\ref{eq2}), but now $\tau_{{\rm CSM}_{\rm clumps}}$ represents the optical depth 
along a clump diameter.  Since the clumps occupy only a small fraction of the volume, they have 
a volume filling factor $\phi \ll 1$.  Additionally, the probability that the line-of-sight 
from a given clump intersects another clump is low, so that a relation between the size of a clump, 
the number density of clumps, and the radial coordinate can be written as  

\begin{equation}
\label{eq13}
\eta~\pi r^2 ~R ~\approx~ N ~<~ 1 ,
\end{equation}

\noindent where $\eta$ is the volume number density of clumps, $r$ is the radius of a clump, 
$R$ is the distance from the center of the SN
to the blastwave-CSM interaction region, and $N$ is the average number of clumps along the 
line-of-sight, with $N$ appreciably lower than unity by definition.  Finally, it is easy to verify 
that there is a relation between the volume filling factor $\phi$, $r$, $R$ and $N$, of the form

\begin{equation}
\label{eq14}
 \phi ~=~ \frac{4}{3} ~ \frac{r}{R} ~N .
\end{equation}

\noindent We can then express the effective optical depth $<\tau_{{\rm eff}}^{0.5}>$ as

\begin{equation}
\label{eq15}
<\tau_{{\rm eff}}^{0.5}> = {\frac{\sqrt{2}}{3}} ~ \tau_{{\rm CSM}_{\rm clumps}}^{0.5} 
\phi^{0.5} N^{0.5} ,
\end{equation}

\noindent where, for initial estimates, we shall take $N \sim 0.5$ and a constant ratio 
$r R^{-1} \sim 0.33$, so that, from equation (\ref{eq14}), $\phi \sim 0.22$.
Before we can fully estimate the mass-loss rate, $\dot M$, for the progenitor
star of SN 1988Z before explosion, we must first fit our parameterized radio
light curve model to the radio dataset for the SN.

\section{Radio Light Curve Description}

Examination of Figure 1 and Table 1 shows that the multi-frequency radio data are best described by 
two time intervals: ``early'' data, which extends roughly from explosion through day 1479, 
and ``late'' data, which roughly extends from day 2129 through the end of the dataset.  
Figure 1 shows clear steepening of the light curves sometime between these two measurement 
epochs, but the actual ``break'' date is somewhat arbitrary, due to uncertainties in the 
flux density measurements for this relatively faint source and the likely smoothness of 
any transition region.  We have therefore chosen to describe the flux density evolution 
separately for these two time intervals, with the period from day $\sim 1500$ to day $\sim 2000$ 
as a transition.  With the explosion date assumed to be 1998 December 1 from optical 
estimates (Stathakis \& Sadler 1991), and extensive trial fitting showing no evidence for 
distant, homogeneous thermal absorption ($K_4 = 0$), for SSA ($K_5=0$), for mixed free-free 
absorption/nonthermal emission ($K_6=0$), or for a local homogeneous thermal absorption component 
($K_2=0$), the model fitting process needs to determine only five parameters ($K_1$, $\alpha$, 
$\beta$, $K_3$, and $\delta^\prime$).  Applying our fitting procedures separately to the early 
and late periods, with the data points between day 1500 and day 2000 included in both fits to 
provide a smooth transition, we find that both the spectral index, $\alpha$, and the clumpy 
absorption parameters, K$_3$ and $\delta^\prime$, are the same in the two time intervals, to within 
the fitting errors.  However, the emission decay rate parameter, $\beta$, steepens significantly,
from $\sim-1.2$ for the early period to $\sim-2.7$ for the late period.  The results of this model
fitting are listed in Table 3.  Note that $K_1$ is a flux density scaling factor, so the 
fact that it has very different values for the early and the late periods is not physically 
significant.

For a purely clumpy CSM ($K_2=0$), the sharp steepening of $\beta$ around day 1750, while 
$K_3$ and $\delta^\prime$ remain unchanged, implies that: 1) the number density of clumps per 
unit volume, $\eta$, starts decreasing more rapidly with radius by approximately $R^{-1.5/m}$ 
(i.e., $\eta_{{\rm day}>1750} = \eta_{{\rm day}<1750}{\left(\frac{R}{R_{{\rm day} 1750}}\right)^{-1.5/m}}$),
with the average characteristics of the individual clumps remaining unchanged, and 2) most of 
the absorption occurs within the emitting clumps themselves.  In other words, the spatial 
distribution is so sparse that the average number of clumps along the line-of-sight is less 
than one ($N < 1$).  This second condition is consistent with Case 3
from Weiler et al.~(2001, 2002) 
and is the basis for our selection of Case 3 for SN 1988Z.  It is
perhaps significant that Weiler et al.~(2001, 2002) found that Case 3 also applies to the radio light curves for the unusual 
SN 1998bw.

Considering other wavelengths, we note that the H$\alpha$ light curve
from Aretxaga et al.~(1999; see their Figure 2) deviates significantly 
from their models
after day $\sim$1000. To investigate this further, we have plotted in
Figure 3 the H$\alpha$ light curve after day 300, the interval for
which we have radio data.  Examination of Figure 3 shows that the
H$\alpha$ light curve is consistent with a slope
steepening from $\beta{_1}=-1.84$ to $\beta{_2}=-2.94$ after day $\sim$1250.  
The H$\alpha$ emission is proportional to $\rho^2$, while for
``typical'' SNe the radio emission is proportional to $\rho^{1.4}$
(i.e., if $\beta$ is the H$\alpha$ light curve decline rate, 
$\beta{_1} = -1.84$ should be 
the same as $\sim 2/1.4 \times \beta_{\rm radio}=1.42\times\beta_{\rm early}=-1.73$,
where $\beta_{\rm early}=-1.22$ from Table 3).
Thus, very good agreement is found between the behavior of the decline for
both the early optical H$\alpha$ emission-line and early radio light curves, to within the
uncertainties.  Also, the change in the slope ($\Delta\beta$) at both 
H$\alpha$ and in the radio should be the same, because the break is due to 
a change in the number of clumps, not in the properties of the clumps and,
therefore, should be wavelength independent:
$\Delta\beta_{{\rm H}\alpha}=(-2.94)-(-1.84)=-1.10$;
$\Delta\beta_{\rm radio}=(-2.73)-(-1.22)=-1.51$, which is also in fair agreement,
considering the uncertainties.  Thus, it appears that the steepening in the
decline for the nonthermal radio light curves is consistent with the 
steepening of the thermal H$\alpha$ light curve.  

\subsection{The Mass-Loss Rate for the SN Progenitor}

We can now use equations (\ref{eq12}) and (\ref{eq15}) to estimate from the 
radio absorption a mass-loss rate for the 
SN 1988Z progenitor star.  With the assumptions for the blastwave and CSM properties 
discussed in \S 3 and the results for the best-fit parameters listed in Table 3, 
we obtain an estimated presupernova mass-loss rate,
$\dot M = 1.2 \times 10^{-4}\ M_\odot$ yr$^{-1}$.  

This high mass-loss rate for SN 1988Z is only appropriate for the last
$\sim10,000$ years before explosion (see \S 5).  At earlier epochs of the progenitor's evolution
the mass-loss rate was considerably lower, as indicated by the much
more steeply declining $\beta$ with unchanged $K_3$ and
$\delta^\prime$ after day $\sim$1750. 

\section{Discussion}

Figure 1, Table 3, and $\S4$ show that the radio light curves for SN 1988Z can be 
described by standard RSN models (Weiler et al.~1986, 1990, 2001; Montes et al.~1997), 
with only one fitting parameter varying with time --- the index of the time evolution 
of the radio emission, $\beta$.  At an age of $\sim$1750 days $\beta$ clearly steepened 
from $\sim-1.2$ to $\sim-2.7$.  This change in radio emission characteristic, $\beta$,
was {\it not\/} accompanied by a corresponding change in the absorption characteristics,
$K_3$ and $\delta^\prime$.  Thus, the change in the radio light curves for SN 1988Z 
cannot be described as only due to a change in the average CSM density, which so effectively 
described the changes in radio evolution for SN 1980K (Montes et al.~1998) and SN 1979C 
(Montes et al.~2000). 

For SN 1988Z the observations are best interpreted as a decrease in the number 
density of clumps in the wind-established CSM after day $\sim1750$, but with all clumps 
having essentially the same average internal characteristics as at earlier times.  Since the shock 
velocity for SN 1988Z is assumed to be $\sim20,000$ \kms ~and the pre-supernova wind velocity 
for an RSG is typically $\sim10$ \kms, an interval of 1,750 days after explosion implies 
that this change took place in the presupernova stellar wind $\sim9,600$ years before explosion. 
With an estimated mass-loss rate in the interval prior to explosion of 
$\sim1.2\times 10^{-4}\ M_\odot$ yr$^{-1}$, it follows that during this last 10,000 
years, the progenitor star shed $\sim1.1~M_\odot$ in a massive ``superwind.''  The abrupt 
steepening of $\beta$ around day 1750 implies that the mass-loss rate was appreciably lower 
before that time, decreasing by about a factor of four by day 4,438.  Even such a decreasing 
mass-loss rate, however, still accounts for an additional $\sim0.8\ M_\odot$ lost in the 
stellar wind by our last measurement on day 4,438 ($\sim25,000$ years before explosion, using 
the same wind and blastwave velocity assumptions).  Since a relatively high mass-loss rate 
must have been maintained over a much longer time, such as the $\sim10^5$ years appropriate 
for a massive star's evolution, an additional several $M_\odot$ must 
have been lost earlier so that the presupernova mass loss for SN 1988Z over the entire 
RSG progenitor lifetime is expected to have been $>2\ M_\odot$, and perhaps much greater.

As discussed in $\S1$, Chugai \& Danziger (1994) studied SN 1988Z in detail and developed 
a model from which they concluded that the ejecta mass is small ($M_{\rm ej} < 1\ M_\odot$),
and that SN 1988Z, therefore, must have originated from a relatively low-mass star, 
$M_{\rm ZAMS} \sim 8$--10 ${M_\odot}$, in sharp contrast with the high-mass (20--40 $M_\odot$) 
progenitor proposed by Van Dyk et al.~(1993a) and Stathakis \& Sadler (1991).  However, it should 
be noted that Chugai \& Danziger derived a very high mass-loss rate for the SN 1988Z progenitor,
$7 \times 10^{-4}\ M_\odot$~yr$^{-1}$, yielding what they call their ``wind density parameter'' 
(our $\dot M/w_{\rm wind}$) of $5 \times 10^{16} ~{\rm g ~cm^{-1}}$.  We believe that our mass-loss 
rate derived from the direct measurement of f-f absorption in the CSM, 
$1.2 \times 10^{-4}\ M_\odot$~yr$^{-1}$ (wind density parameter, 
$\dot M/w_{\rm wind} = 7.3 \times 10^{15} ~{\rm g ~cm^{-1}}$), is likely more appropriate.  
Since the critical factors in the modeling by Chugai \& Danziger are the product of the ejecta 
mass and wind density parameter related to the energy of the explosion, for a given explosion 
energy our 6--7 times smaller wind density parameter implies, for their model, a factor of 
6--7 times greater ejecta mass, $M_{\rm ej} \sim 5$--10 $M_\odot$, rather than their estimate of 
$M_{\rm ej} \le 1\ M_\odot$.  Even our value for the ejecta mass may still be an underestimate because, 
by the Chugai \& Danziger model, it represents only the mass that was observed to interact with 
the pre-supernova wind and may not involve the entire ejecta mass.  With an ejecta mass of as 
much as $\sim 10 ~M_\odot$ and a high radio luminosity, which places SN 1988Z among the brightest 
radio supernovae, we conclude that the stellar progenitor of SN 1988Z must have been a very massive star, 
perhaps as high as $\sim20$ -- 30~$M_\odot$, as previously suggested by Van Dyk et al.~(1993a).

It is worth noting that all of the well-studied cases of ultra-bright
RSNe (say, having $\ge$10 times the radio luminosity of SN 1979C at 6 cm peak), such as SNe 1988Z,
1986J, and 1998bw, provide evidence for a highly-clumped CSM with 
presupernova wind mass-loss rates up to $\sim10^{-4} M_\odot$ yr$^{-1}$ and unusually high 
shock velocities $\gtrsim20,000$ ~\kms.  From optical and ultraviolet 
spectroscopy Fransson et al.~(2002) also find for the SN IIn 1995N evidence for high
shock velocities, a ``superwind'' phase for the progenitor characterized by a 
high mass-loss rate prior to explosion, and a clumpy CSM.
Such high-rate mass loss, clumping of the CSM, 
and consequent high blastwave velocities 
may be the signature of particularly massive star progenitors (see also the
discussion in Fransson et al.~2002).
It is yet to be determined how the progenitors of SNe IIn, which must retain
at least some of
their hydrogen envelopes until the end of their lifetimes, and the progenitors of possibly extreme
Type Ib/c SNe, such as SN 1998bw, which must have lost their entire hydrogen
envelopes before explosion, are related, if at all.  

\section{Conclusions}

The radio emission from SN 1988Z followed an evolution well described by standard models
up to an age of $\sim1750$ days ($\sim4.8$ yrs), after which its behavior evolved into a much more 
rapid decline in radio emission without a corresponding change in
radio absorption parameters.  This is the first
time we have witnessed this in any well-studied RSN, and it implies that highly radio-luminous RSNe,
such as SN 1988Z and, possibly, SN 1986J, may follow a different evolutionary path in their 
presupernova mass-loss than that for more ``normal'' RSNe, such as SNe 1979C and 1980K.  
However, it is interesting to note that, if one assumes a blastwave velocity $\sim2,000$ times faster 
than the RSG wind-established CSM ($\sim20,000$ \kms\ vs.~$\sim10$ \kms), the timescale for 
such a variation in SN 1988Z of $\sim10^4$ yr, is similar to that seen
for SNe 1979C and 1980K (see, e.g., 
for SN 1979C, Montes et al.~1998; for SN 1980K, Montes et al.~2000), and also for SN 1998bw 
(Weiler et al.~2001).  

In addition, further analysis of the results from Chugai \& Danziger (1994),
which implied an extremely high mass-loss rate of $7 \times 10^{-4} M_\odot$ yr$^{-1}$ and a
relatively low-mass progenitor with $M_{\rm ZAMS} \sim 8$--10 $M_\odot$ for SN 1988Z, indicates that their estimates are probably 
unrealistic.  
Using their modeling, our lower mass-loss rate, $1.2 \times 10^{-4}\ M_\odot$ yr$^{-1}$, implies 
a significantly higher ejecta mass and, therefore, a higher ZAMS mass of $\sim20$--30 $M_\odot$
for the progenitor star.

We have noted that the H${\alpha}$ data from
Aretxaga et al.~(1999), which are also indicative of the
SN shock-CSM interaction (Chevalier \& Fransson 1994), also begin to deviate significantly from their
model by day $\ga$1250.  Although a detailed comparison is beyond the
scope of this paper, both the slope before the break and the
magnitude of the break are roughly consistent with the radio results.

From our analysis we propose that SNe, such as SNe 1988Z, 1986J, and
1998bw (possibly the counterpart to GRB 980425), with possible very massive star
progenitors ($M_{\rm ZAMS} > 20\ M_\odot$) and associated
massive winds ($\dot {M} \ga 10^{-4}\ M_\odot$~yr$^{-1}$), have very highly-clumped, wind-established 
CSM and unusually high blastwave velocities ($>20,000$~\kms).  

\acknowledgments

KWW thanks the Office of Naval Research (ONR) for the 6.1
funding supporting this research.  This work was primarily done while CLW was
at Montgomery Blair High School, Silver Spring, MD, serving as a summer
student at NRL.
CLW thanks the NRL SEAP and STEP programs for supporting his summer 
work.  Additional information and data on radio supernovae can be found on {\it http://rsd-www.nrl.navy.mil/7214/weiler/} and linked pages.

\begin{deluxetable}{lcccccc}
\tablewidth{4.8in}
\tabletypesize{\footnotesize}
\tablecaption{Flux Density Measurements for SN 1988Z\tablenotemark{a}}
\tablehead{
\colhead{Obs.~Date} & \colhead{Age} & \colhead{VLA} & \colhead{$S$ (20 cm)} & \colhead{$S$ (6 cm)} 
& \colhead{$S$ (3.6 cm)} & \colhead{$S$ (2.0 cm)}\\
\colhead{(UT)} & \colhead{(days)} & \colhead{Config.} & \colhead{(mJy)} & \colhead {(mJy)} 
& \colhead{(mJy)} & \colhead{(mJy)}} 
\startdata
1988 Dec 1  & $\equiv$ 0 & & & & \\
1989 Dec 21 &  385 & D   & \nodata        & $0.67\pm 0.07$ & \nodata        & \nodata \\
1990 Feb 12 &  438 & DnA & \nodata        & $0.77\pm 0.09$ & \nodata        & \nodata \\
1990 May 29 &  544 & A   & \nodata        & $1.21\pm 0.09$ & \nodata        & \nodata \\
1990 Jul 12 &  588 & BnA & \nodata        & $1.26\pm 0.09$ & $2.10\pm 0.14$ & \nodata \\
1990 Sep 7  &  645 & B   & $<$0.39        & $1.38\pm 0.09$ & $1.72\pm 0.14$ & $1.15\pm 0.23$ \\
1990 Dec 14 &  743 & C   & $<$0.48        & $1.78\pm 0.10$ & $2.09\pm 0.13$ & $1.43\pm 0.19$ \\
1991 Jun 11 &  922 & A   & $0.47\pm 0.07$ & $1.90\pm 0.11$ & $1.68\pm 0.10$ & $1.31\pm 0.25$ \\
1991 Sep 17 & 1020 & A   & \nodata        & $1.85\pm 0.10$ & $1.48\pm 0.09$ & $0.68\pm 0.16$ \\
1992 Jan 26 & 1151 & CnB & $1.04\pm 0.13$ & $1.67\pm 0.10$ & $1.50\pm 0.09$ & $1.17\pm 0.18$ \\
1992 Oct 13 & 1412 & A   & $0.91\pm 0.08$ & $1.57\pm 0.09$ & $1.22\pm 0.10$ & $0.91\pm 0.22$ \\
1992 Dec 19 & 1479 & A   & $1.52\pm 0.09$ & $1.72\pm 0.11$ & $1.04\pm 0.08$ & $0.78\pm 0.19$ \\
1993 May 4  & 1615 & B   & $1.16\pm 0.10$ & $1.21\pm 0.09$ & $0.98\pm 0.06$ & $0.57\pm 0.15$ \\
1993 Aug 23 & 1726 & BnA & $1.05\pm 0.15$ & $1.28\pm 0.09$ & \nodata        & \nodata \\
1993 Dec 5  & 1830 & D   & \nodata        & \nodata        & $0.87\pm 0.07$ & \nodata \\
1994 Feb 18 & 1905 & DnA & $1.09\pm 0.16$ & $0.98\pm 0.17$ & $0.86\pm 0.14$ & \nodata \\
1994 May 4  & 1980 & BnA & $1.42\pm 0.16$ & $0.97\pm 0.13$ & $0.52\pm 0.08$ & \nodata \\
1994 Sep 30 & 2129 & CnB & $1.06\pm 0.15$ & $0.91\pm 0.05$ & $0.56\pm 0.08$ & \nodata \\
1995 May 19 & 2360 & D   & \nodata        & $0.62\pm 0.05$ & $0.42\pm 0.04$ & $0.26\pm 0.09$ \\
1995 Oct 3  & 2497 & BnA & $0.78\pm 0.07$ & $0.40\pm 0.04$ & $0.23\pm 0.04$ & $<$0.38 \\
1996 Oct 5  & 2865 & DnA & $0.72\pm 0.15$ & $0.36\pm 0.09$ & $0.32\pm 0.06$ & $<$0.36 \\
1997 Jan 23 & 2975 & BnA & $0.68\pm 0.11$ & $0.30\pm 0.08$ & \nodata        & \nodata \\
1998 Feb 10 & 3358 & DnA & \nodata        & \nodata        & $<0.15$        & \nodata \\
1998 Feb 13 & 3361 & DnA & $0.35\pm 0.06$ & $0.30\pm 0.05$ & \nodata        & \nodata \\
1998 Oct 19 & 3609 & B   & $0.39\pm 0.07$ & $0.20\pm 0.03$ & \nodata        & \nodata \\
1999 Jun 13 & 3846 & DnA & $0.19\pm 0.06$ & $<$0.20        & \nodata        & \nodata \\
1999 Oct 1  & 3956 & DnA & $0.22\pm 0.06$ & $0.17\pm 0.04$ & \nodata        & \nodata \\
2001 Oct 19 & 4340 & A   & $0.19\pm 0.03$ & $<$0.18        & $<$0.18        & \nodata \\
2001 Jan 25 & 4438 & DnA & $0.17\pm 0.06$ & $<$0.26        & $<$0.27        & \nodata \\
\enddata
\tablenotetext{a}{All upper limits are 3$\sigma$.}
\end{deluxetable}

\clearpage

\begin{deluxetable}{lcccc}
\tablecolumns{5}
\tablewidth{3.5in}
\tabletypesize{\footnotesize}
\tablecaption{Flux Density Measurements for Secondary Calibrators}
\tablehead{
\colhead{Obs.~Date} & \colhead{$S$ (20 cm)} & \colhead{$S$ (6 cm)} & \colhead{$S$ (3.6 cm)} 
& \colhead{$S$ (2.0 cm)}\\
\colhead{(UT)} & \colhead{(Jy)} & \colhead{(Jy)}  & \colhead{(Jy)} & \colhead{(Jy)}} 
\startdata
\cutinhead{J1051+213} \nl
1989 Dec 21 & \nodata & 0.921   & \nodata & \nodata \\
1990 Feb 12 & \nodata & 0.885   & \nodata & \nodata \\
1990 May 29 & \nodata & 0.901   & \nodata & \nodata \\
1990 Jul 12 & \nodata & 0.907   & 0.894   & \nodata \\
1990 Sep 7  & 1.092   & 1.002   & 0.942   & 1.123   \\
1990 Dec 14 & 1.118   & 0.948   & 1.069   & 0.967   \\
1991 Jun 11 & 0.962   & 0.997   & 1.162   & 1.203   \\
1991 Sep 17 & \nodata & 1.091   & 1.358   & 1.336   \\
\cutinhead{J1125+261} \nl
1992 Jan 26 & 0.992   & 1.069   & 1.001   & 0.806   \\
1992 Oct 13 & 0.948   & 1.102   & 0.954   & 0.839   \\
1992 Dec 19 & 0.937   & 1.068   & 0.926   & 0.729   \\
1993 May 4  & 0.967   & 1.008   & 0.871   & 0.703   \\
1993 Aug 23 & 0.892   & 0.986   & \nodata & \nodata \\
1993 Dec 5  & \nodata & \nodata & 0.887   & \nodata \\
1994 Feb 18 & 0.910   & 1.049   & 0.934   & \nodata \\
1994 May 4  & 0.920   & 1.056   & 0.891   & \nodata \\
1994 Sep 30 & 0.943   & 1.082   & 0.984   & \nodata \\
1995 May 19 & \nodata & 1.088   & 0.936   & 0.749   \\
1995 Oct 3  & 0.880   & 1.098   & 0.966   & 0.781   \\
1996 Oct 5  & 0.899   & 1.059   & 0.854   & 0.657   \\
1997 Jan 23 & 0.893   & 1.060   & \nodata & \nodata \\
1998 Feb 10 & \nodata & \nodata & 0.918   & \nodata \\
1998 Feb 13 & 0.961   & 0.911   & \nodata & \nodata \\
1998 Oct 19 & 0.989   & 0.967   & \nodata & \nodata \\
1999 Jun 13 & 0.957   & 0.927   & \nodata & \nodata \\
1999 Oct 1  & 0.937   & 0.988   & \nodata & \nodata \\
2000 Oct 19 & 0.924   & 0.887   & 0.709   & \nodata \\
2001 Jan 25 & 0.933   & 1.070   & 0.936   & \nodata \\
\enddata
\end{deluxetable}


\begin{deluxetable}{cccc}
\tablecolumns{4}
\tablewidth{4.5in}
\tablecaption{Fitting Parameters for SNe 1988Z} 
\tablehead{
\colhead{Parameter} & \multicolumn{3}{c}{SN 1988Z\tablenotemark{a}} \\
\colhead{} & \multicolumn{3}{c}{\hrulefill} \\
\colhead{} & \colhead{Early} & \colhead{Global} &\colhead{Late} \\
\colhead{} & \colhead{${\rm day}<2000$} & \colhead{} & \colhead{${\rm day}>1500$}}
\startdata
$K_1$ & $1.2 \times 10^4$ & & $9.1 \times 10^8$  \\
$\alpha$ & & {$-0.72$} & \\
$\beta$  & $-1.22$ & & $-2.73$ \\
$K_3$ & & {$3.19 \times 10^8$} & \\
$\delta^{\prime}$ & & {$-2.87$} & \\
$t_0$\tablenotemark{b} & & {$\equiv$ 1988 Dec 01} & \\
${\rm L_{6\ cm\ peak}}$ (erg s$^{-1}$ Hz$^{-1}$) & & {$2.3 \times 10^{28}$} & \\
$t = (t_{\rm 6\ cm\ peak} - t_0)$ (days) & & {911} & \\ 
$\dot M$ ($M_\odot$ yr$^{-1}$)\tablenotemark{c} & & { $1.2 \times 10^{-4}$} & \\
\enddata
\tablenotetext{a}{Because of a change in the evolution of the radio emission from SN 1988Z during 
the interval between day $\sim$1500 and day $\sim$2000, the data are split into two overlapping 
intervals of ``early,'' from day 385 through day 1980, and ``late,'' from day 1615
through day 4438.}
\tablenotetext{b}{Adopted from Stathakis \& Sadler~(1991).}
\tablenotetext{c}{Assuming $w_{\rm wind}= 10~{\rm km}~{\rm s}^{-1}$, 
$v_{\rm i}=v_{\rm blastwave} = 20,000~\rm{km}~\rm{s}^{-1}$, 
$\rm{T}=20,000~\rm{K}$, ${\rm t_i} = 45~{\rm days}$, $N = 0.5$,  
$r R^{-1} = 0.33$, and $\phi = 0.22$ and calculating $m = 0.83$ and $<\tau_{{\rm eff}}^{0.5}> = 0.16$.}
\end{deluxetable}

\clearpage

\begin{figure}
\figurenum{1}
\epsscale{0.75}
\caption{Radio ``light curves'' for SN 1988Z in MCG
+03-28-022.  The four wavelengths, 2 cm (14.9 GHz, 
{\it open circles}, {\it solid line}), 3.6 cm (8.44 GHz, 
{\it crosses}, {\it dashed line}), 6 cm (4.87 GHz, {\it open
squares}, {\it dot-dash line}), and 20 cm (1.48 GHz, {\it open 
triangles}, {\it dotted line}), are shown together with their best-fit 
model light curves.  The SN age is in days since the adopted
explosion date of 1988 December 1 (Stathakis \&
Sadler 1991).  Because the decline index $\beta$ of the radio emission 
steepened between, roughly, day $\sim$1500 and day
$\sim$2000, an iterative fitting procedure was used (see \S4), and a
break in the model radio light curves can be seen near day 1750.}
\end{figure}


\begin{figure}
\figurenum{2}
\epsscale{0.60}
\caption{Flux density measurements of the secondary
calibrators J1051+213 and J1125+261 at 2 cm (14.9 GHz, {\it open circles}), 
3.6 cm (8.44 GHz, {\it crosses}), 6 cm (4.87 GHz, {\it open
squares}), and 20 cm (1.48 GHz, {\it open triangles}).  For clarity the data
points at each frequency are offset from one another by adding 1.0 Jy
to the 2 cm values, 0.5 Jy to the 3.6 cm values, 
and $-$0.5 Jy to the 20 cm values.  The epoch at which the
secondary calibrator was changed from J1051+213 to J1125+261 is marked.}
\end{figure}


\begin{figure}
\figurenum{3}
\epsscale{0.75}
\plotone{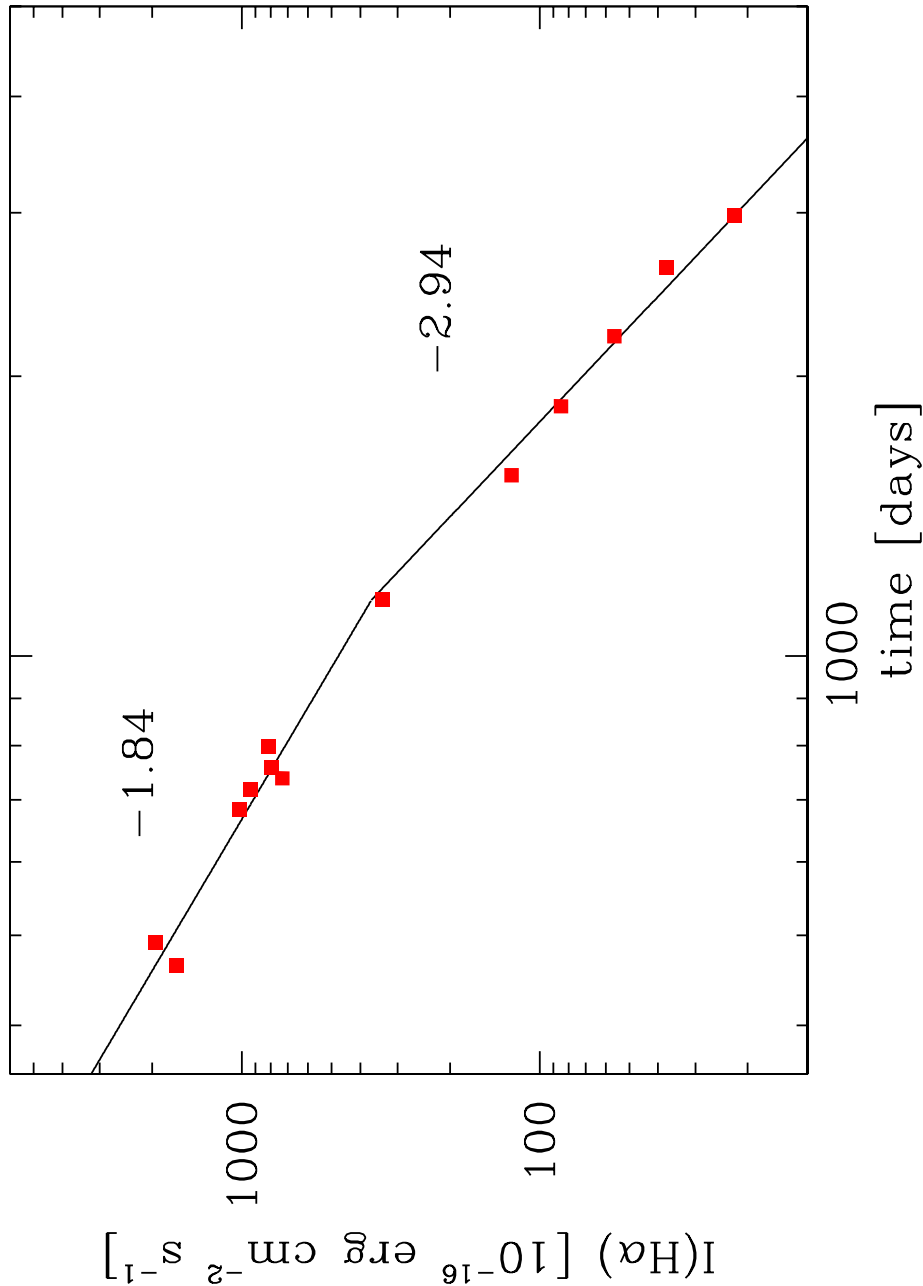}
\caption{Evolution of the H$\alpha$ line intensity, from
Aretxaga et al.~(1999), during the same time range when radio
observations were obtained, $\sim$300 to $\sim$3000 days after
explosion.  Note a likely steepening of the decline rate, $\beta$, 
for the H$\alpha$ intensity
after $\sim1250$ days, with $\beta_1=-1.84$ before the break and 
$\beta_2=-2.94$ thereafter.}
\end{figure}


\begin{thebibliography}{}
\bibitem[Aretxaga et al.~(1999)]{Aretxaga99} Aretxaga, I., Benetti, S., Terlevich, R.\ J., Fabian, A.\ C., Cappellaro, E., Turatto, M., \& della Valle, M.\ 1999, \mnras, 309, 343
\bibitem[Bregman \& Pildis (1992)]{Bregman92} Bregman, J.\ N., \& Pildis, R.\ A. 1992, \apjl, 398, L107
\bibitem[Cappellaro, Turatto, \& Candeo (1988)]{Cappellaro88} Cappellaro, E., Turatto, M., \& Candeo, G. 1988, \iaucirc, 4691
\bibitem[Chevalier (1982a)]{Chevalier82a} Chevalier, R.\ A. 1982a, \apj, 259, 302
\bibitem[Chevalier (1982b)]{Chevalier82b} Chevalier, R.\ A. 1982b, \apjl, 259, L85
\bibitem[Chevalier \& Fransson (1994)]{Chevalier94} Chevalier, R.\ A. \& Fransson, C. 1994, \apj, 420, 268
\bibitem[Chugai \& Danziger (1994)]{Chugai94} Chugai, N.\ N. \& Danziger, I.\ J. 1994, \mnras, 268, 173
\bibitem[Fabian \& Terlevich (1996)]{Fabian96} Fabian, A.\ C. \& Terlevich, R. 1996, \mnras, 280, L5
\bibitem[Filippenko (1989)]{Filippenko89} Filippenko, A.\ V. 1989, \iaucirc, 4713
\bibitem[Filippenko (1991a)]{Filippenko91a} Filippenko, A.\ V. 1991a, in Supernovae: The Tenth Santa Cruz Workshop in Astronomy and Astrophysics. Ed., S.E. Woosley (Springer-Verlag, New York), 467
\bibitem[Filippenko (1991b)]{Filippenko91b} Filippenko, A.\ V. 1991b, in SN 1987A and Other Supernovae. Eds., I.\ J. Danziger and K. Kjar (ESO, Garching bei Muenchen), 343
\bibitem[Filippenko, Matheson, \& Barth (1994)]{Filippenko94} Filippenko, A., Matheson, T., \& Barth, A.\ 1994, \aj, 108, 222
\bibitem[Fransson et al.~(2002)]{Fransson02} Fransson, C., Chevalier, R.\ A.,
Filippenko, A.\ V., Leibundgut, B., Barth, A.\ J., Fesen, R.\ A., Kirshner, 
R.\ P., Leonard, D.\ C., Li, W.\ D., Lundqvist, P., Sollerman, J., \& Van Dyk,
S.\ D. 2002, \apj, 572, 350
\bibitem[Heathcote et al.~(1988)]{Heathcote88} Heathcote, S., Cowley, A., \& Hartwick, D. 1988, \iaucirc, 4693
\bibitem[Houck et al.~(1998)]{Houck98} Houck, J.\ C., Bregman, J.\ N., Chevalier, R.\ A., Tomisaka, K. 1998, \apj, 493, 431
\bibitem[Iwamoto et al.~(1998)]{Iwamoto98} Iwamoto, K. et al.~1998, \nat, 395, 672
\bibitem[Kirshner, Leibundgut, \& Smith (1989)]{Kirshner89} Kirshner, R., Leibundgut, B., \& Smith, C. 1989, \iaucirc, 4900
\bibitem[Leibundgut et al.~(1991)]{Leibundgut91} Leibundgut, B., Kirshner, R.\ P., Pinto, P.\ A., Rupen, M.\ P., Smith, R.\ C., Gunn, J.\ E., \& Schneider, D.\ P. 1991, \apj, 372, 531
\bibitem[Montes, Weiler, \& Panagia (1997)]{Montes97} Montes, M.\ J., Weiler, K.\ W., \& Panagia, N.\ 1997, \apj, 488, 792 
\bibitem[Montes et al.~(1998)]{Montes98} Montes, M.\ J., Van Dyk, S.\ D., Weiler, K.\ W., Sramek, R.\ A., \& Panagia, N.\ 1998, \apj, 506, 874 
\bibitem[Montes et al.~(2000)]{Montes00} Montes, M.\ J., Weiler, K.\ W., Van Dyk, S.\ D., Panagia, N., Lacey, C.\ K., Sramek, R.\ A., \& Park, R. 2000, \apj, 532, 1124
\bibitem[Natta \& Panagia (1984)]{Natta84} Natta, A., \& Panagia, N. 1984, \apj, 287, 228
\bibitem[Osterbrock (1974)]{Osterbrock74} Osterbrock, D.\ E.\ 1974, Astrophysics of Gaseous Nebulae (San Francisco; Freeman), pp.\ 82-87
\bibitem[Pastorello et al.~(2002)]{Pastorello02} Pastorello, A., Turatto, M.,
Benetti, S., Cappellaro, E., Danziger, I.\ J., Mazzali, P.\ A., Patat, F., 
Filippenko, A.\ V., Schlegel, D.\ J., \& Matheson T. 2002, \mnras, 333, 27
\bibitem[Pollas (1988)]{Pollas88} Pollas, C. 1988, \iaucirc, 4691
\bibitem[Rupen et al.~(1987)]{Rupen87} Rupen, M.\ P., van Gorkom, J.\ H., Knapp, G.\ R., Gunn, J.\ E., \& Schneider, D.\ P. 1987, \aj, 94, 61
\bibitem[Schlegel (1990)]{Schlegel90} Schlegel, E.\ M. 1990, \mnras, 244, 269
\bibitem[Spyromilio (1994)]{Spyromilio94} Spyromilio, J.\ 1994, \mnras, 266, 61
\bibitem[Sramek, Panagia, \& Weiler (1984)]{Sramek84} Sramek, R.\ A., Panagia, N., \& Weiler, K.\ W.\ 1984, \apjl, 285, L59 
\bibitem[Sramek, Weiler, \& Panagia (1990)]{Sramek90a} Sramek, R.\ A., Weiler, K.\ W., \& Panagia, N. 1990, \iaucirc, 5112
\bibitem[Stathakis \& Sadler (1991)]{Stathakis91} Stathakis, R.\ A. \& Sadler, E.\ M. 1991, \mnras, 250, 786
\bibitem[Turatto et al.~(1993)]{Turatto93} Turatto, M., Cappellaro, E., Danziger, I.\ J., Benetti, S., Gouiffes, C., \& della Valle, M. 1993, \mnras, 262, 128
\bibitem[Van Dyk et al.~(1993a)]{VanDyk93a} Van Dyk, S., Sramek, R.\ A., Weiler, K., \& Panagia, N.\ 1993a, \apjl, 419, 69
\bibitem[Van Dyk et al.~(1993b)]{VanDyk93b} Van Dyk, S.~D., Sramek, R.~A., Weiler, K.~W., \& Panagia, N.\ 1993b, \apj, 409, 162 
\bibitem[Van Dyk et al.~(1994)]{VanDyk94} Van Dyk, S., Weiler, K., Sramek, R., Rupen, M., \& Panagia, N.\ 1994, \apjl, 432, 115
\bibitem[Weiler et al.~(1986)]{Weiler86} Weiler, K.\ W., Sramek, R.\ A., Panagia, N., van der Hulst, J.\ M., \& Salvati, M. 1986, \apj, 301, 790
\bibitem[Weiler, Panagia, \& Sramek (1990)]{Weiler90} Weiler, K.\ W., Panagia, N., \& Sramek, R.\ A. 1990, \apj, 364, 611
\bibitem[Weiler et al.~(1991)]{Weiler91} Weiler, K.\ W., Van Dyk, S.\ D., Discenna, J.\ L., Panagia, N., \& Sramek, R.\ A. 1991, \apj, 380, 161
\bibitem[Weiler et al.~(1992a)]{Weiler92a} Weiler, K.\ W., Van Dyk, S.\ D., Pringle, J., \& Panagia, N.\ 1992a, \apj, 399, 672
\bibitem[Weiler et al.~(1992b)]{Weiler92b} Weiler, K.\ W., Van Dyk, S.\ D., Panagia, N., \& Sramek, R.\ A.\ 1992b, \apj, 398, 248
\bibitem[Weiler, Panagia, \& Montes (2001)]{Weiler01} Weiler, K.\ W., Panagia, N., \& Montes, M. 2001, \apj, 562, 670
\bibitem[Weiler et al.~(2002)]{Weiler02} Weiler, K.\ W., Panagia, N., Montes, M.\ J., \& Sramek, R.\ A. 2002, \araa, 40, 387 
\end{thebibliography}
\end{document}